\documentclass[%
reprint,
 amsmath,amssymb,
 aps,
prb,
]{revtex4-1}

\usepackage{mhchem}
\usepackage{braket}
\usepackage{graphicx}
\usepackage{dcolumn}
\usepackage{bm}

\DeclareMathOperator{\sign}{sign}

\newcommand{\Xbar}{\overline{ \mathrm{X}}}
\newcommand{\kp}{ \mathbf{k} \cdot \mathbf{p} }

\begin{document}

\title{Topological states on uneven (Pb,Sn)Se (001) surfaces}

\author{Rafa\l{} Rechci\'nski}
 \email{rafmr@ifpan.edu.pl}

\author{Ryszard Buczko}%
 \email{buczko@ifpan.edu.pl}
\affiliation{%
 Institute of Physics, Polish Academy of Sciences, Aleja Lotnikow 32/46, PL-02668 Warsaw, Poland
}%

\date{\today}

\begin{abstract}
The impact of surface morphology on electronic structure of topological crystalline insulators is studied theoretically. As an example, the structure of topologically protected electronic states on a (001) (Pb,Sn)Se surface with terraces of atomic height is modeled. Within the envelope function model  it is shown that valley mixing, the phenomenon responsible for the peculiar "double Dirac cone" shape of the surface state dispersion, depends crucially on the structure of the surface. By varying the width and the number of atomic layers in the terraces, a comprehensive explanation of recent experimental findings, i.e., the emergence of 1D states bound to odd-height atomic step edges as well as the collapse of "double Dirac cone" structure on a rough surface, is achieved. This approach allows us also to determine topological indices characterizing terraces and their interfaces.  In the (001) surface of (Pb,Sn)Se the adjacent terraces turn out to be described by different values of the winding number topological invariant.
\end{abstract}

\maketitle

\section{Introduction}

In 2012 it was shown that some IV-VI compounds and their crystalline solid solutions, such as: SnTe, (Pb,Sn)Se
and (Pb,Sn)Te, belong to a newly discovered class of topological matter. In these so called topological crystalline insulators (TCI)\cite{Hsieh2012,Dziawa2012,Tanaka2012,Xu2012} the nontrivial topology of electronic bands is protected by $\{110\}$ mirror planes. In the topological phase the band gap in the four L points of the Brillouin zone (BZ) of these rock salt crystals has to be inverted. This is always the case in SnTe, however, in the solid solutions the sign of the band gap can be tuned between normal and TCI phases by temperature, and Sn content, or pressure. This was demonstrated in (Pb,Sn)Se by ARPES,\cite{Wojek2014} and infrared measurements.\cite{Xi2014}

At the TCI surfaces, and also at the interfaces between the TCI and a normal insulator (NI), spin-polarized states of massless electrons appear due to the bulk-boundary correspondence.\cite{Ando2013a,Bansil2016} Volkov and Pankratov predicted such states already in 1985, however, without linking them to nontrivial topology of the bulk bands.\cite{Volkov1985}

\begin{figure}
\includegraphics{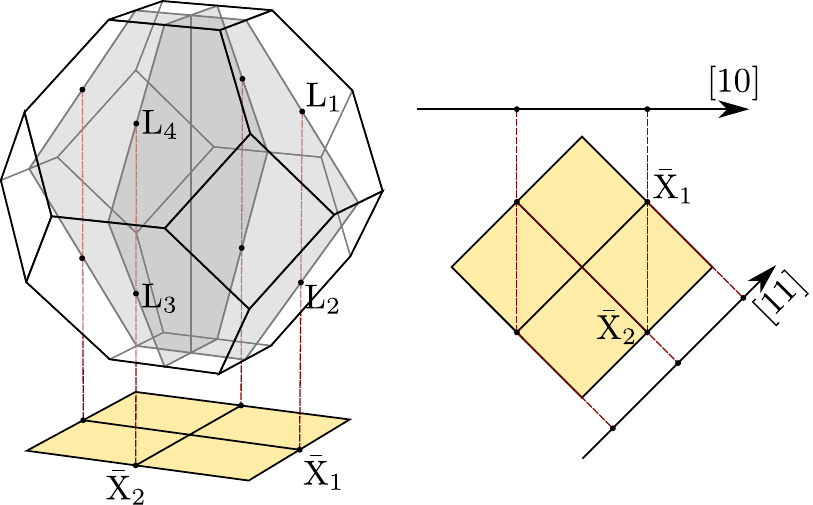}
\caption{\label{fig:brillouin}  Left: Bulk Brillouin zone of rock-salt structure and its projection onto the (001) surface Brillouin zone. Right: (001) surface Brillouin zone and its projections onto 1D Brillouin zones of [10] and [11] step edges.}
\end{figure}

Depending on the surface orientation, L points are projected either into different points of the two dimensional (2D) BZ or in pairs.\cite{Liu2013,Safaei2013} In the first case, e.g, for $\{111\}$ surface, the topological states are described by Dirac cone dispersions, and the Dirac points are located at L point projections in the 2D BZ of the surface. The second case occurs only for $\{nnm\}$ surfaces with $n$ and $m$ of opposite parity.\cite{Safaei2013} There the number of surface states at the projection of the two L points (denoted $\Xbar$ for $\{001\}$ surfaces) doubles. Due to hybridization resulting from mixing of L valleys, the dispersion features two separated in energy Dirac points at $\Xbar$ and two secondary Dirac points in the middle of the gap, which are shifted away from $\Xbar$ along the mirror symmetry line. Only in the case of $\{001\}$ cleavage surfaces two symmetry lines exist (see Fig.~\ref{fig:brillouin}) and protect two pairs of secondary Dirac points. The valley splittings at $\Xbar$ observed by ARPES are significant -- they are of the order, however always lower, than the bulk band gap. The dispersion of surface states in the vicinity of $\Xbar$ is depicted in Fig.~\ref{fig:disp}.

Lately it has been shown by ARPES that valey splitting can be substantially reduced by small terraces of a normal insulator deposited on the top of TCI surface.\cite{Polley2018} The tight binding description of (Pb,Sn)Se (001) surface overgrown with PbSe in Ref.~\onlinecite{Polley2018} demonstrates that the splitting oscillates with the height of terraces. Maximal value is attained for a flat surface or terraces of the height of an even number of monolayers (even-height), i.e. an integer number of lattice constants $a_0$. The splitting reduces to zero in the case of odd-height terraces described by half-integer multiples of $a_0$. This phenomenon can be related to the $2 \pi/a_0$ distance between the two interacting L valleys and is explained further in this paper. A similar effect of valley splitting reduction has been described in the case of Si nanostructures with disorder or steps at the interfaces.\cite{Ando1982}

In this paper we continue the study of the valley splitting of surface states in the presence of atomic steps but now in a (Pb,Sn)Se homostructure. For this purpose we derive an appropriate and simple model based on the envelope function (EF) approximation. By comparing the results of the model with the results of tight binding (TB) method we find that it provides physically grounded and quantitatively adequate description of the surface states. With a proper choice of parameters the model can be applied to surfaces of other TCIs in the SnTe class and their planar heterostructures. 

We show that in consistency with the previously studied case small odd-height terraces can reduce the splitting to zero. The splitting can be recovered, however, in the presence of very wide terraces, typical on a cleavage surface. In this case, results of our model show that the surface and odd-height terraces define domains of different topology characterized by opposite winding numbers. As a consequence, at the steps which form the domain boundaries we can find zero-energy one dimensional states of similar origin as edge states of graphene ribbons.\cite{Ryu2002} The step states were recently discovered by Sessi et al.\cite{Sessi2016} on the surface of (Pb,Sn)Se with scanning tunneling spectroscopy and described by a toy model and TB approximation. Our description allows deeper understanding of their properties and topological origin.

\section{\label{sec:ema}The model}
\subsection{$\mathbf{k} \cdot \mathbf{p}$ model for a flat surface }
\begin{figure}
\includegraphics[scale=1]{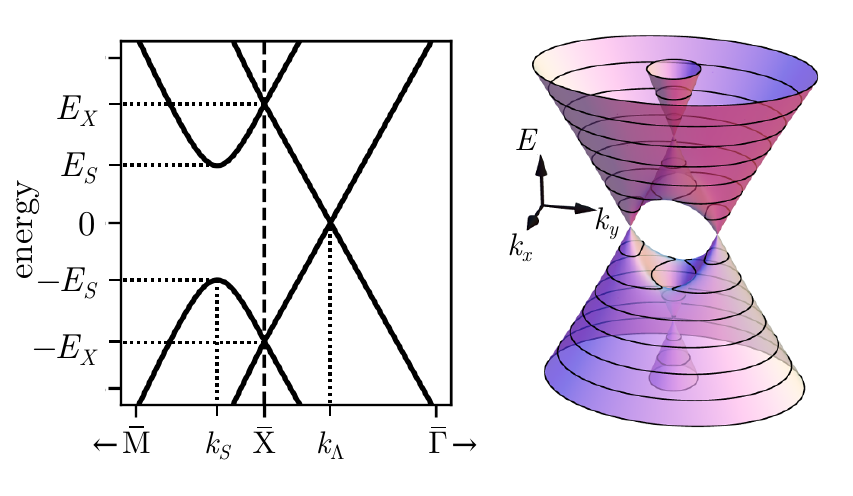}
\caption{The energy dispersion given by the Hamiltonian~\eqref{eq:kpHamiltonian}: the cross section along high-symmetry lines (left) and a 3D plot of the dispersion in the vicinity of $\Xbar$ (right).} \label{fig:disp}
\end{figure}

A simple description of topological states on a flat (001) surface of a IV-VI TCI is provided by the $\mathbf{k} \cdot \mathbf{p}$ Hamiltonian analogous to the one in Ref.~\onlinecite{Liu2013}
\begin{equation} \label{eq:kpHamiltonian}
H( \mathbf{k} ) = m \tau_x + k_x (v_x s_y + v''_x \tau_z s_z) - k_y v_y s_x ,
\end{equation}
where $\mathbf{k}=0$ is the $\Xbar$ point. The basis of Pauli matrices $\tau$ are states arising from the $\mathrm{L}_1$ ($\tau_z = 1$) and the $\mathrm{L}_2$ ($\tau_z = -1$) valleys. Pauli matrices $s$ operate between the Kramers partners within each of the valleys. The valley mixing is described by $m$ and $v''_x$ terms. In their absence the Hamiltonian describes a doubly degenerate Dirac cone. The dispersion of~\eqref{eq:kpHamiltonian} is presented in Fig.~\ref{fig:disp}. The two protected crossings are located at $ \mathbf{k}_\Lambda = (0, \pm |m|/v_y)$. The saddle points at $\mathbf{k}_S = (\pm |m v_x|/(v_x^2 + v''^2_x)$ have energies $E_S = \pm |m v''_x|/\sqrt{v_x^2 + v''^2_x}$. Energies of crossings at $\Xbar$ are $E_X = \pm m$. Following the notation of Ref.~\onlinecite{Liu2013}, the symmetries of the Hamiltonian are denoted by:
\begin{equation} \label{eq:symm_ops}
M_x = -i s_x, \quad M_y = -i \tau_x s_y, \quad \Theta = i s_y K,
\end{equation}
where $M_x$ and $M_y$ are $(1 \bar{1} 0)$ and $(110)$ mirror reflections, and $\Theta$ is the time reversal operator. While these symmetries allow more terms in~\eqref{eq:kpHamiltonian}, our comparison to TB calculations of (Pb,Sn)Se (001) surface states justifies restricting the Hamiltonian to just the four terms featured in the formula above. 

It is important to note, that there is a gauge freedom in defining $H( \mathbf{k} )$. Rotating the Hamiltonian by $U_\theta=\exp( i \theta \tau_z s_x )$ with any angle $\theta$ changes the form of~\eqref{eq:kpHamiltonian}, while leaving the operators~\eqref{eq:symm_ops} invariant.\footnote{Supplemental Material of Ref. \onlinecite{Liu2013}} In this paper we choose a gauge in which $m\tau_x$ is the only $k$-independent term.

Furthermore, we recognize that the Hamiltonian exhibits chiral symmetry, i.e. $\Gamma H \Gamma = -H$, where $\Gamma = \tau_y s_z$. Consequently, any closed contour in the two-dimensional $\mathbf{k}$ space of the crystal surface can be characterized by the 1D winding number $\nu$ topological invariant.\cite{Chiu2016} This provides a useful tool for identifying topologically protected edge states.\cite{Ryu2002} Details of calculations of $\nu$ in our system are available in S1 and S2 of the Supplemental Material.\footnote{See Supplemental Material for details on the calculation of the winding number, derivation of the correspondence between $\kp$ and TB models, proof of existence of the zero energy modes, and comparison of EF and TB results.}

\subsection{Atomic steps in the envelope function model } \label{subs:steps_model}
To include the steps on the surface into the EF calculation we first present a simplified reasoning which we later verify by comparison to the empirical TB model. 

We will assume that every basis state of the $\tau$ subspace of~\eqref{eq:kpHamiltonian}, denoted as $F_{\mathrm{L}_i}(\mathbf{r})$ ($i$ denotes the $\mathrm{L}$ valley, the spin index is omitted), can be expressed as a solution of an effective mass equation for one $\mathrm{L}$ valley only.\cite{Volkov1985} For surface states on a flat $(001)$ face, we can write:\footnote{Supplemental Material of Ref.~\onlinecite{Polley2018}.}
\begin{equation} \label{eq:general_Phi1Phi2}
\begin{split}
F_{\mathrm{L}_1}(\mathbf{r}) = f(z) \Psi_{\mathrm{L}_1}( \mathbf{r} ),\\
F_{\mathrm{L}_2}(\mathbf{r}) = f^*(z) \Psi_{\mathrm{L}_2}( \mathbf{r} ),
\end{split}
\end{equation}
where $\Psi_{\mathrm{L}_i}$ is a Bloch wavefunction from the $\mathrm{L}_i$ valley.  The complex envelope functions $f$ is exponentially decaying for $z \rightarrow + \infty$ (bulk region). The Bloch wavefunctions consist of the edges of bulk conduction and valence bands.\cite{Volkov1985}

We will consider states $F'_{\mathrm{L}_i}$ identical to the ones in Eq.~\eqref{eq:general_Phi1Phi2}, but anchored to a surface that is displaced with respect to the original one by one atomic monolayer. This is achieved by shifting the states by vector $\mathbf{t}_3 = a_0/2 (1,0,1)$, corresponding to removing one atomic layer from the surface. To satisfy the condition
\begin{equation}
F_{\mathrm{L}_i}(\mathbf{r}-\mathbf{t}_3) = F'_{\mathrm{L}_i}(\mathbf{r})
\end{equation}
we calculate
\begin{equation} \label{eq:translacja_Blocha}
\Psi_{\mathrm{L}_i} (\mathbf{r}-\mathbf{t}_3) = e^{ - i \mathbf{L }_i \cdot \mathbf{t}_3 } \Psi_{\mathrm{L}_i} (\mathbf{r})
\end{equation}
and
\begin{equation}  \label{eq:przesuniete_stany}
\begin{split}
F'_{\mathrm{L}_1}(\mathbf{r}) = - f\left(z- \frac{a_0}{2} \right) \Psi_{\mathrm{L}_1}( \mathbf{r} )\\
 F'_{\mathrm{L}_2}(\mathbf{r}) = f^*\left(z- \frac{a_0}{2} \right)  \Psi_{\mathrm{L}_2}( \mathbf{r} ).
\end{split}
\end{equation}
Assuming that $f$ is varying slowly within the distance of $a_0/2$ we can approximate $
F'_{\mathrm{L}_1}(\mathbf{r}) \approx - F_{\mathrm{L}_1}(\mathbf{r}), F'_{\mathrm{L}_2}(\mathbf{r}) \approx  F_{\mathrm{L}_2}(\mathbf{r})$. This allows us to denote the operator of translation by $\mathbf{t}_3$ as 
\begin{equation} \label{eq:translation}
T_{\mathbf{t}_3} \approx - \tau_z.
\end{equation}
We note that $- \tau_z$ is gauge invariant with respect to unitary transformation $U_\theta$.

Let us now consider two vast terraces (labelled $A$ and $B$) on the (001) face, separated by a step edge of single atomic height . While the surface states far from the terrace edge are described by the same Hamiltonian matrix \eqref{eq:kpHamiltonian}, the basis states for these matrices should be $F_{\mathrm{L}_i}$ on one terrace and $F'_{\mathrm{L}_i}$ on the other. To express the Hamiltonian of the full system in one basis we take $H^{A} = H$ and $H^{B} = (- \tau_z) H (- \tau_z)$. Equivalently $H^{B}$ can be obtained from $H^{A}$ by substituting $m \rightarrow -m$. We arrive at an interesting result that even though the states on each of the terraces are the same, due to their relative displacement it is justified to describe them by two different Hamiltonians with an inverted order of states.

Note that $T_{n \mathbf{t}_3} = (- \tau_z)^n$, which equals $1$ for even $n$ and $- \tau_z$ for odd $n$. This means that the surface states divide into two classes: the states which occupy terraces with even number of layers, and the states on terraces with odd number of layers. 

\subsection{Correspondence with a realistic tight binding model}
The appealing result~\eqref{eq:translation} requires verification in a more realistic model which does not rely on the coarse description~\eqref{eq:general_Phi1Phi2} of basis states. Therefore, we perform calculations of the (001) surface states in 18-orbital sp$^3$d$^5$ TB approximation with spin-orbit interactions included. We choose parameters describing \ce{Pb_{0.68}Sn_{0.32}Se} at temperature $100\,\mathrm{K}$, derived according to the procedure described in Ref.~\onlinecite{Wojek2013}. The basis states of~\eqref{eq:kpHamiltonian} can now be expressed as superpositions of the four numerical TB eigenvectors $\ket{\Phi_i ( \mathbf{k}_{\Xbar_1}  )}$ ($i$ numbers the states) describing surface states at $\Xbar_1 = \pi/a_0 (1,1,0)$. 

A detailed description of the procedure of finding these superpositions is available in S4 of the Supplemental Material.\cite{Note2} Here we present an abridged explanation, excluding details not essential to our argument. 

The basis of~\eqref{eq:kpHamiltonian} is defined by eigenspaces of diagonal operators $\tau_z$ and $s_z$. To find the eigenstates of the former we analyze $\ket{\Phi_i ( \mathbf{k}_{\Xbar_1}  )}$ Fourier transformed along the $[001]$ axis. In this way we evaluate contributions of quasimomenta $\mathbf{k} = (\pi/a_0,\pi/a_0,k_\perp)$ into the states, including the vicinities of $ \mathrm{L}_1 = \pi/a_0 (1,1,1)$ and $\mathrm{L}_2 = \pi/a_0 (1,1,-1)$, i.e. the $\mathrm{L}$ valleys. In the Fourier basis we can project our states onto subspaces $k_\perp \in [0,2 \pi/a_0]$ and $k_\perp \in [-2 \pi/a_0,0]$. This allows us to find linear combinations of $\ket{\Phi_i }$ that have maximal contribution from $\mathrm{L}_1$ and minimal from $\mathrm{L}_2$ and vice versa. These new states we assign as eigenstates of $\tau_z$.

Since we expect the operator $T_{\mathbf{t}_3}$ to be gauge invariant and to not mix the $s$ degree of freedom, the evaluation of the eigenstates of $s_z$ is not relevant here. Nevertheless we perform it as a check of consistency. The calculation is outlined in S4 of the Supplemental Material.\cite{Note2}

\begin{figure}
\includegraphics[scale=1]{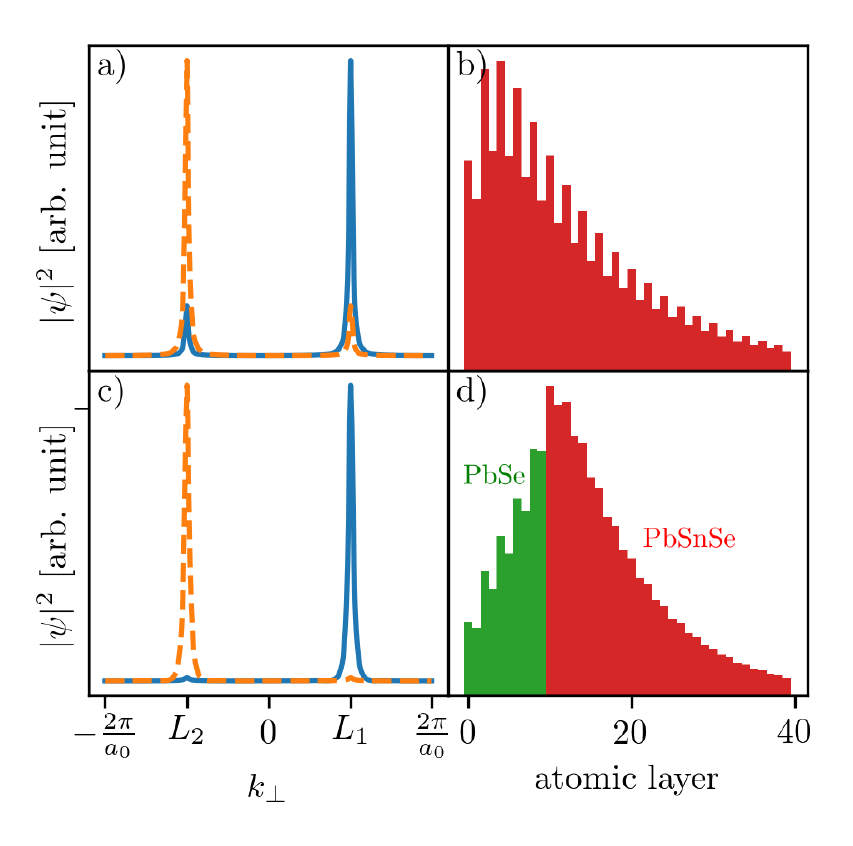}
\caption{The valley-resolved basis states of Hamiltonian~\eqref{eq:kpHamiltonian} calculated from the TB results for Pb$_{0.68}$Sn$_{0.68}$Se. Panels (a) and (c) show squared moduli of Fourier transforms along the axis perpendicular to the surface, while (b) and (d) depict squared moduli of wavefunction amplitudes at the first 40 layers near the surface. Subplots (a) and (b) show the case of a clean (Pb,Sn)Se surface. Out of the four basis states two are composed mostly of states from the $\mathrm{L}_1$ valley (solid blue line) and other two (dashed orange line) mostly of states from the $\mathrm{L}_2$ valley, while all four have the same real-space probability density (b). Subplots (c) and (d) show the same analysis in the case of valley-splitting decreased by putting 10 layers of PbSe on the (Pb,Sn)Se surface. Colors of the bars denote different materials. } \label{fig:states_basis}
\end{figure}

Finally, we obtain four TB wavefunctions $\ket{ F^s_{\mathrm{L}_i}  }$ corresponding to~\eqref{eq:general_Phi1Phi2}. Their probability densities are shown in Fig.~\ref{fig:states_basis}(a,b).

Now we can check whether relation~\eqref{eq:translation} applies also to states $\ket{ F^s_{\mathrm{L}_i}  }$. We compute the overlap matrix $\braket{F^{s}_{\mathrm{L}_i}| T_{\mathbf{t}_3}|F^{s'}_{\mathrm{L}_j}}$ with $T_{\mathbf{t}_3}$ defined in the TB basis. We find that every $\ket{ F^s_{\mathrm{L}_i}  }$ is orthogonal to all shifted basis states except for its counterpart $T_{\mathbf{t}_3} \ket{ F^s_{\mathrm{L}_i}  }$. The overlap matrix has the form $- \gamma \tau_z $. For our choice of \ce{Pb_{0.68}Sn_{0.32}Se} $\gamma = 0.7$.

The overlap does not attain the absolute value of $1$, firstly because one of the states occupies a space that comprises one atomic layer more, but also because the assumption that the basis states are  constrained to separate $\mathrm{L}$ valleys is not very well satisfied, as seen in Fig.~\ref{fig:states_basis}(a). This discrepancy is a consequence of the large valley splitting pushing the surface states close to edges of bulk conduction and valence bands. The orbital makeup of each of the surface states originates mostly from the nearer bulk band. The fact that a valley-split surface state may be an unequal mixture of the Bloch wavefunctions describing the bulk band edges constitutes a parameter which is not considered in the basic EF model, where the basis of $\tau$ subspace is taken as states derived strictly from one valley, unperturbed by scattering to the other valley. Such states, in absence of mixing, would land in the middle of the gap and would, therefore, consist equally of the conduction and the valence band orbitals. This is not the case for the TB result for a free surface. Calculation with valley splitting decreased by deposition of 10 layers of NI PbSe on the (Pb,Sn)Se surface is shown as an example in Fig.~\ref{fig:states_basis}(c,d). There the split states are much closer to the middle of the gap and have similar orbital makeup. Therefore it is possible to almost perfectly separate the basis states into individual valleys. For that case $\gamma = 0.96$.

We conclude that approximation~\eqref{eq:translation} more accurately describes systems in which valley splitting is small compared to the bulk band gap. However, it remains valid also in the case of a free (Pb,Sn)Se surface. This is further supported by results obtained from our model showing very good agreement with TB calculations (see the comparison in Fig.~\ref{fig:evolution}(e,f) and S5 of the Supplemental Material\cite{Note2}). In the subsequent EF calculations we use parameters fitted to the dispersion of the surface states of \ce{Pb_{0.68}Sn_{0.32}Se} at $100\,\mathrm{K}$ temperature, that is $m=36\,\mathrm{meV}$, $v_x = 0.24\,\mathrm{eV \cdot nm}$, $v''_x =0.13 \,\mathrm{eV \cdot nm}$, $v_y = 0.22 \,\mathrm{eV \cdot nm}$. At this temperature $a_0 = 6.062\,\textrm{\AA}$.

\section{Results}

\subsection{States localized at odd-height step edges}
First, we consider the steps consisting of an odd number of layers (odd-height steps). In our model such a step edge between terraces is a sharp interface between two half planes, one described by $H^A$, and the other by $H^B$. For infinitely long terraces the quasimomentum $k_\parallel$ parallel to the step remains a good quantum number. This allows us to treat each $H^{A(B)}(k_\parallel = \mathrm{const.})$ as a separate 1D problem.

We recall that the toy model in Ref.~\onlinecite{Sessi2016} predicted states localized at the odd-height step edges. These states had flat dispersion and existed only at quasimomenta contained between the two $\mathbf{k}_\Lambda$ points. To determine the possible topological origin of the states at the interface we calculate the winding number $\nu$ associated with $H^{A(B)}(k_\parallel = \mathrm{const.})$, describing 1D cuts through the 2D $k$-space.\cite{Ryu2002}

We find that $\nu = -\sign{(m v''_x)}$ for any section of the $k$-plane that goes between the two $\mathbf{k}_\Lambda$ points, and $\nu = 0$ for any other section, as shown of Fig.~\ref{fig:winding}(a). See S2 of the Supplemental Material for details of the calculations.\cite{Note2} As the sign of $m$ changes at step edges, while $v''_x$ remains constant, at such a step edge the value of the winding number $\nu$ between the Dirac points, i.e., for $|k_\parallel| < \mathbf{k}_\Lambda \cdot \hat{ \mathbf{e}}_\parallel$, changes by $\Delta \nu = \pm 2$. Such value of $\Delta \nu$ indicates that the interface can host two localized states. Because of chiral symmetry those states must have opposite energies $E$ and $-E$. 

\begin{figure}
\includegraphics[scale=1.]{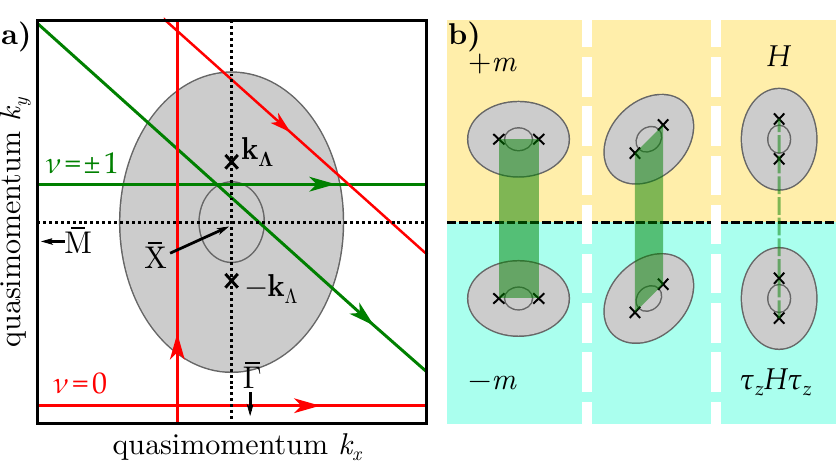}
\caption{a) The winding numbers of 1D cuts through $H(k_x,k_y)$. In grey a schematic top view of the dispersion from Fig.~\ref{fig:disp} is shown. The Dirac points are marked by crosses. Lines passing between the Dirac points (green) have $\nu = \pm 1$, while others (red) have $\nu =0$. b) Selected alignments of the $k_x$ and $k_y$ axes with respect to the step edge. Highlighted in green are quasimomenta parallel to the step edge at which the adjacent terraces have opposite $\nu$ (with a limiting case shown rightmost). } \label{fig:winding}
\end{figure}

To find the explicit form of the states we use the EF approximation. In Hamiltonian~\eqref{eq:kpHamiltonian} we substitute
\begin{equation}
\begin{split}
k_\perp \rightarrow -i \frac{d}{dx_\perp}, \\ m \rightarrow m \sign{ x_\perp },
\end{split}
\end{equation}
where $x_\perp$ points perpendicular to the step edge. We find that any two eigenfunctions of such a Hamiltonian localized around $x_\perp=0$ must necessarily belong to the same chiral subspace, associated with one of the projectors $P_\pm = \frac{1}{2}(1 \pm \Gamma)$ (which one exactly depends on the sign of $\Delta \nu$).

We encounter an uncommon result, that even though there are two chiral states on the same edge, both must have $E=0$, i.e., zero bias with respect to the energy of the Dirac crossings at $\mathbf{k}_\Lambda$. The proof can be found in S3 of the Supplemental Material.\cite{Note2}

\subsubsection{Steps in the $[11]$ direction}
The vicinity of $\Xbar_1 = \pi/a_0 (1,1,0)$ point can be treated by setting $x_\parallel = y, x_\perp = x$. Without loss of generality we will assume that $v''_x>0$ and that $m(x)$ has a negative sign for $x<0$ and a positive sign for $x>0$. We can now restrict our search for solutions to the $P_-$ chiral subspace. In the basis which diagonalizes $\Gamma$, obtained by rotation of the original basis by $U = \frac{1}{\sqrt{2}} (\tau_z + \tau_y)$, the EF equation for zero-energy modes becomes
\begin{equation}
-i h_x \frac{d \psi }{d x} + h_0(x) \psi + h_y k_y \psi = 0,
\end{equation}
where
\begin{equation}
\begin{split}
h_0(x) &= - (m \sign{x})  \sigma_x, \quad (m>0) \\
h_x &= - i v_x \sigma_z - i v''_x  \sigma_x, \\
 h_y &= -v_y \sigma_0. \\
\end{split}
\end{equation}

Looking for the solutions in the form $\psi(x) = e^{\lambda x} f_0$ we find two modes on the $x>0$ half plane
\begin{equation} \label{eq:modes}
\begin{split}
\lambda_-^\pm =  - \frac{ m  v''_x}{w_x^2} \pm \frac{ \sqrt{ v_y^2 k_y^2  w_x^2 - m^2 v_x^2 } }{w_x^2},
\end{split}
\end{equation}
where $w_x^2 = v_x^2 + v''^2_x$. The corresponding unnormalized $f_0$ vectors are
\begin{equation} \label{eq:vectors}
f_-^\pm = 
\begin{pmatrix}
m + v''_x \lambda_-^\pm \\
-v_y k_y - v_x \lambda_-^\pm \\
\end{pmatrix}.
\end{equation}
Modes on the $x<0$ half plane can be obtained by switching $m\rightarrow -m$, to find
\begin{equation}
\begin{split}
\lambda_+^\pm =  + \frac{ m  v''_x}{w_x^2} \pm \frac{ \sqrt{ v_y^2 k_y^2  w_x^2 - m^2 v_x^2 } }{w_x^2},
\end{split}
\end{equation}
\begin{equation} 
f_+^\pm = 
\begin{pmatrix}
-m + v''_x \lambda_+^\pm \\
-v_y k_y - v_x \lambda_+^\pm \\
\end{pmatrix}.
\end{equation}

Depending on the value of $k_y$ the formulas~\eqref{eq:modes} can describe two (for $|k_y|<k_\Lambda$) or one (for $|k_y|>k_\Lambda$) evanescent mode on each half-plane. In the first case the continuity condition is satisfied trivially, as the two linearly independent vectors~\eqref{eq:vectors} span the entire Hilbert space, so there is always a superposition of $f_-^\pm$ from the $x>0$ half plane that will match any mode on the $x<0$ half plane.\footnote{The degenerate point $k_y = \pm \frac{m v_x}{w_x v_y}$, where $f_-^+=f_-^-$,  has to be treated separately, but also allows two different modes localized on the step edge } In the second case it can be proven by inspection that the vectors $f^-_-$ and $f^+_+$ associated to evanescent modes at $x>0$ and $x<0$, respectively, are linearly independent.

We conclude, therefore, that it is possible to create two continuous EFs that are localized at the step edge for $k_y$ between the two protected crossings near $\Xbar_1$. 

\begin{figure*}
\includegraphics[scale=1]{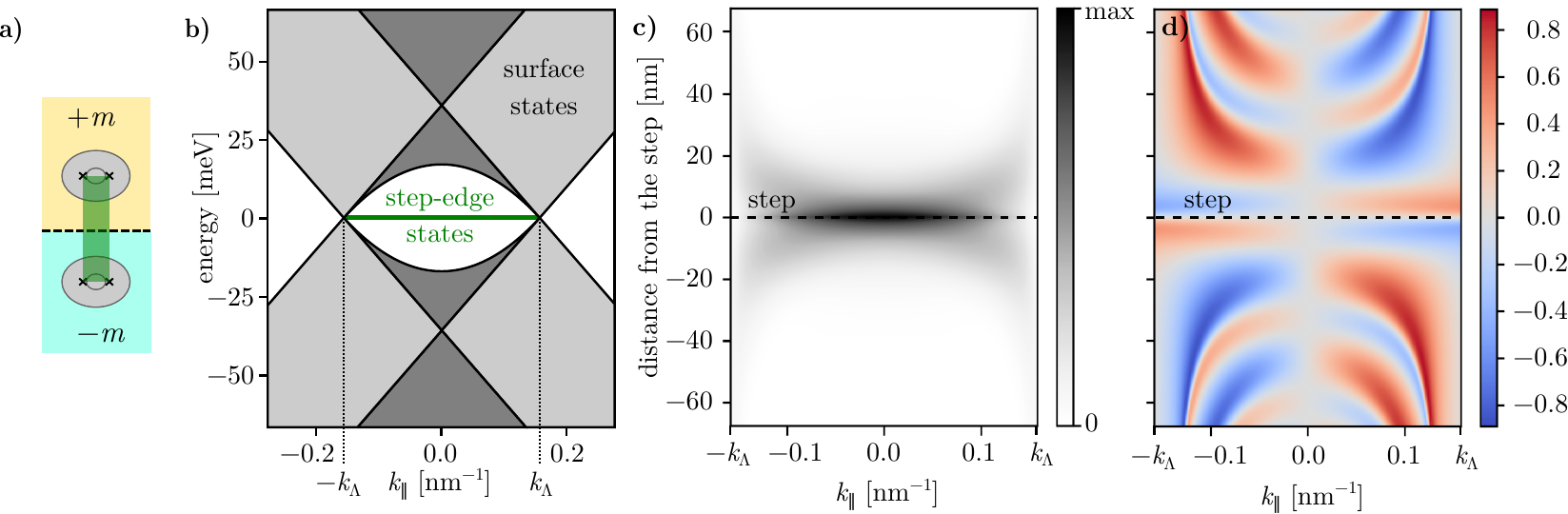}
\caption{ Solution of the EF equation showing states localized at a single odd-height $[11]$ step edge. a) Alingment of the surface states spectrum with respect to the step, b) the flat dispersion of the localized state shown on background of other surface states, c) sum of square moduli of the two orthogonal solutions, d) summed degree of $s_z$ polarization of both states.  } \label{fig:zero_mode}
\end{figure*}

For $k_y=0$ there are two elegant orthogonal solutions of the EF equation
\begin{equation}
\psi^\pm (x) = N e^{ (-v_x'' \pm i |v_x| ) \frac{|m|  }{v_x^2 + v''^2_x} |x|  }  \begin{pmatrix}
\pm i \\
1 \\
\end{pmatrix}
\end{equation}
where $N = \sqrt{ \frac{ |m| v_x'' }{ 2 (v_x^2 + v''^2_x)  }  }$. Note that localization length is inversely proportional to the saddle point energy $E_S$. Figure~\ref{fig:zero_mode} shows the analytical solutions for $|k_y|<| k_\Lambda|$. In addition to the shape of the EF we calculate also the degree of $s_z$ polarization $\Sigma(x, k_y) = [\sum_\psi \psi^\dagger(x, k_y) s_z \psi(x, k_y)] / [ \sum_\psi |\psi(x, k_y)|^2 ]$, which shows away from the step and the symmetric point $\Xbar$ an oscillation of the spin degree of freedom.

To investigate the vicinity of $\Xbar_2 = 2 \pi/a_0 (1,-1,0)$ we set $x_\parallel = -x, x_\perp = y$. In this case we don't find any states localized at step edges, as both Dirac points $\mathbf{k}_\Lambda$ project onto $k_\parallel=0$.

Next, we turn to steps between terraces of finite width. We assume a periodic sequence of alternating terraces that are infinite in the direction parallel to the steps but have finite widths $d_1$ and $d_2$ in the perpendicular direction. In this case the EF equation cannot be easily solved analytically. Let us first rewrite Hamiltonian~\eqref{eq:kpHamiltonian} as
\begin{equation}
H = m \tau_x + \chi_x k_x + \chi_y k_y = m \tau_x + \bm{\chi} \cdot \mathbf{k},
\end{equation}
where $\bm{\chi}$ is a vector of $4 \times 4$ matrices. Now we can write the periodicity condition
\begin{equation} \label{eq:cont_110}
f_0 =  e^{-i \tau_z \chi_\perp^{-1} (H_\parallel-E) \tau_z d_2 }   e^{-i \chi_\perp^{-1} (H_\parallel-E) d_1 } f_0,
\end{equation}
where $f_0$ is the wavefunction at the beginning of one of the terraces, $H_\parallel = m \tau_x + \chi_\parallel k_\parallel$, while $\chi_\perp$ and $\chi_\parallel$ are obtained by an appropriate rotation of $\bm{\chi}$. Equation~\eqref{eq:cont_110} can be solved numerically to find values of $E$ and $k_\parallel$ that admit continuous states on the periodic array of terraces.

\begin{figure*}
\includegraphics[width=\textwidth]{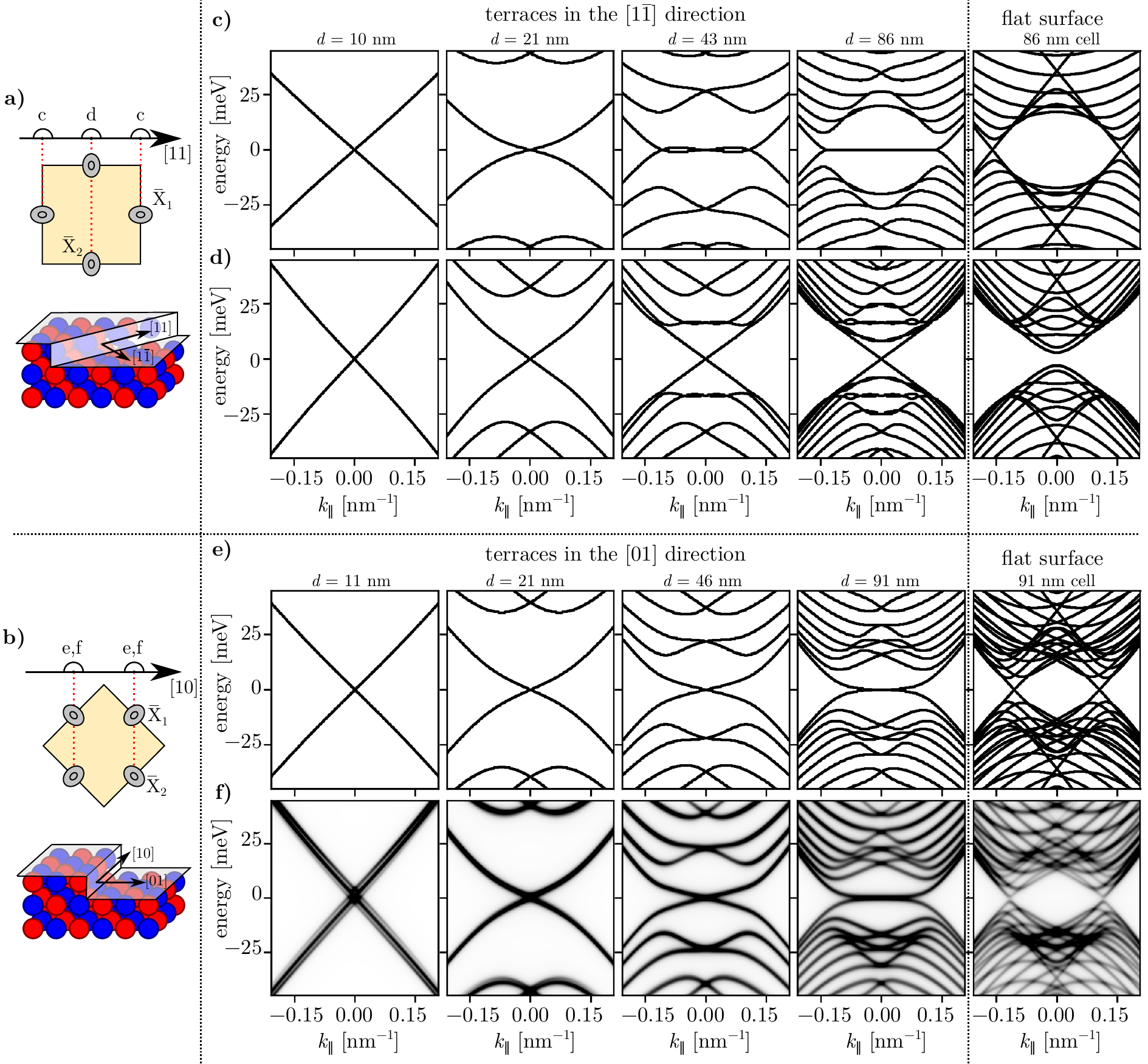}
\caption{Calculated spectra of surface states of \ce{Pb_{0.68}Sn_{0.32}Se} in the presence of evenly distributed, equally wide, parallel terraces of one monolayer height. Above each plot $d=d_1 =d_2$ defines the step-to-step distances. Schematics of the alignment of [11] (a) and [10] (b) steps with respect to the surface Brillouin zone and the crystal structure. Spectra obtained using the EF model for the vicinity of $\Xbar_1$ (c) and $\Xbar_2$ (d) for periodic arrays of [11] steps with different $d$. Spectra obtained in the EF model (e) and the TB model (f) for arrays of [10] steps with different $d$. } \label{fig:evolution}
\end{figure*}

The results for the vicinities of $\Xbar_1$ and $\Xbar_2$, for $d_1 = d_2$, are shown in Figs.~\ref{fig:evolution}(c) and~\ref{fig:evolution}(d), respectively. In the first series of plots we see that the flat states exist only for sufficiently wide terraces (above $30\,\mathrm{nm}$ for the chosen parameters), which is consistent with previously published experimental data.\cite{Sessi2016} For narrower terraces, the spectrum shows a crossing at zero-energy at the projection of $\Xbar_1$ with no traces of the two Dirac crossings at projections of $\mathbf{k}_\Lambda$. This effect we identified as a collapse of the valley splitting of the (001) surface states.\cite{Polley2018} We discuss this case in more detail in a further part of the article. Apart from the flat states we observe that the energy spacings between other states agree with momentum quantization within the width of a single terrace. This is evident upon comparison with an artificially quantized spectrum of a flat surface (rightmost panel of Fig.~\ref{fig:evolution}(c)). Inspection of the EFs obtained from the calculation confirms that indeed each of the terraces hosts a ladder of quantum nanoribbon-like states, which exhibit little leakage to adjacent terraces. Thus a domain structure in which the step edges play the role of domain walls is created.

In the vicinity of $\Xbar_2$, while no flat states are formed, we observe that a zero-energy crossing appears at the projection of $\Xbar_2$. The crossing persists for any terrace width. It's existence for wide terraces can be explained using the winding number argument. A step edge that is slightly tilted away from $[11]$ is required to host two flat zero-energy modes between the projections of $\pm \mathbf{k}_\Lambda$, as for that segment $\Delta \nu = \pm 2$. If we adiabatically return the step edge back to $[11]$ the projections get closer and closer to each other, and the segment of flat states gets shorter, finally merging to one point (as in Fig.~\ref{fig:winding}(b)), which in this limit remains at zero energy. 

Our simple EF model shows very good agreement with the TB approach, while greatly reducing the computational cost. A comparison with TB calculations performed for the same terraces can be found in S5 of the Supplemental Material.\cite{Note2}

\subsubsection{Steps in the $[10]$ direction}
The step edges along the $[10]$ direction are more commonly found on the faces of cleaved crystals than the $[11]$ ones discussed above. As shown in Fig.~\ref{fig:evolution}(b), in this case the two $\Xbar$ points project onto one point in the 1D Brillouin zone of the step edge. $\Xbar_1$ hosts states arising from the $\mathrm{L}_1$ and $\mathrm{L}_2$ valleys, while states at $\Xbar_2$ come from valleys $\mathrm{L}_3$ and $\mathrm{L}_4$. The evaluation of possible scattering between the two momenta is beyond the scope of the simple EF approximation, contrary to the TB method which inherently includes Bloch wavefunctions from the entire $k$-space. Therefore, within the EF model we will limit our analysis to the vicinity of just one of the $\Xbar$ points. This can be interpreted as fully suppressing the scattering. For this direction of the steps we set $x_\parallel = \frac{x+y}{\sqrt{2}}, x_\perp = \frac{x-y}{\sqrt{2}} $.

The calculations are performed for periodic arrays of terraces. From the continuity conditions analogous to~\eqref{eq:cont_110} we derive a series of plots for various widths (with $d_1 = d_2$) of terraces shown in Fig.~\ref{fig:evolution}(e). Much like in the case of $[11]$ steps, the presence of narrow terraces suppresses valley splitting and causes the collapse of the two Dirac cones into one crossing. Also analogously the step edges between wide terraces host states with zero-energy, however, formation of the flat states requires greater terrace widths than in the case of $[11]$ steps which is due to larger localization lengths.

Results of realistic TB calculations performed for the same geometry of terraces exhibit remarkable resemblance to the results of the EF model (compare Figs.~\ref{fig:evolution}(e) and~\ref{fig:evolution}(f)). This leads us to the conclusion that on the step edge there is no significant scattering between states arising from different $\Xbar$ points.

The TB calculations of surface spectral density were obtained using the recursive Green's function method.\cite{LopezSancho1985} Its application to a surface with terraces is described in S5 of the Supplemental Material.\cite{Note2}

\subsection{Rough surface}
In this subsection we present results for states on the $(001)$ (Pb,Sn)Se surface with protruding atomic islands of sizes of tens of lattice constants. Such surface morphology is characteristic for samples grown by molecular beam epitaxy. We believe that the qualitative aspects of the spectral function of such a surface can be judged by considering equations of the form~\eqref{eq:cont_110} with $\chi_\parallel$ corresponding to various directions of the step edge, and setting $d_1$ and $d_2$ of the order of tens of lattice constants or lower. Thus, we model the superficial disorder as a series of densely-spaced parallel atomic ridges. Although this model does not reflect the true geometry of the inhomogeneities on the surface, we use it first of all for the sake of its simplicity. In view of the lack of significant mixing of the two $\Xbar$ valleys shown in the previous section, we believe that such approach is suitable for this problem.

The results for different directions of step edges are shown on the leftmost panels of Fig.~\ref{fig:evolution}. All those cases show degenerate 1D Dirac spectra with the Dirac points located at the projection of $\Xbar$. Since this effect persists for all directions of step edges, we recognize it as the experimentally observed suppression of the valley splitting on the rough surface in (Pb,Sn)Se overgrown with PbSe.\cite{Polley2018} Figure~\ref{fig:evolution} shows only results for evenly spaced step edges, where exactly 50\% of the surface terminates at even layers, and 50\% at odd ones. For different even to odd ratios the collapse of the splitting is partial (not shown here).

\begin{figure}
\includegraphics[scale=1]{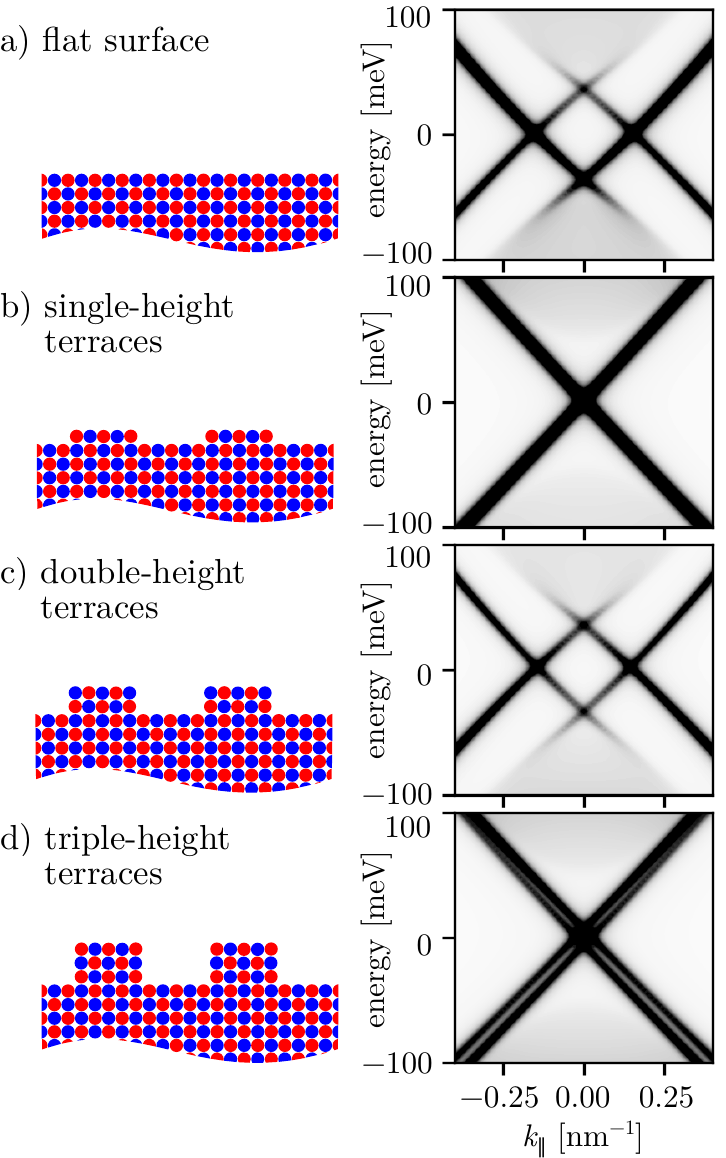}
\caption{Tight binding spectra of surface states of \ce{Pb_{0.68}Sn_{0.32}Se} obtained in presence of $5\,\mathrm{nm}$ wide terraces of various heights.} \label{eq:var_heights}
\end{figure}

For the example of \ce{Pb_{0.68}Sn_{0.32}Se} shown in Fig.~\ref{fig:evolution}, the transition from the disordered regime of the collapsed Dirac cone to the regime of a domain structure of surface states with localized modes on the domain walls happens at around $30$ to $60\,\mathrm{nm}$ distance between step edges, depending on the direction of the edge. It should be noted that on rough surfaces steps of different heights may exist in close proximity, but only steps with heights equal to odd numbers of atomic layers should be taken into account. Steps of double height are mostly invisible to the surface states as evidenced by TB calculation shown in Fig.~\ref{eq:var_heights}. Here the spectral function calculated in presence of such steps is identical to that of a flat surface. The same Figure shows also that steps of triple height produce a slightly weaker suppression of valley splitting than the single height ones, as a result of a higher mismatch of the wavefunctions on adjacent terraces. Nevertheless we can confirm that surface states in the presence of steps of heights 1 and 3 belong to one class, while surface states in the presence of steps of height 2 and 0 (no steps) to the other.

The oscillation of the valley splitting with the terrace height can be understood by recalling that the two interacting $L$ valleys are separated in $\mathbf{k}$ space by distance of $2 \pi/a_0$ perpendicular to the (001) surface. As seen in Eqs.~\eqref{eq:translacja_Blocha} and~\eqref{eq:przesuniete_stany} a translation of the surface states by one monolayer (of height $a_0/2$) introduces a phase difference of $\pi$ between the Bloch states arising from $\mathrm{L}_1$ and $\mathrm{L}_2$ points. Thus, a dense pattern of small single-height terraces kills spatial phase coherence of surface states, and destroys interference between the valleys. A translation by $a_0$ or its integer multiple leaves the relative phases of $\mathrm{L}_1$ and $\mathrm{L}_2$ unchanged. Therefore, an even-height step does not affect the phase coherence and the intervalley interference.

Our results suggest that roughness of the surface can considerably influence functionality of future topological devices, e.g., the topological transistor designed by Liu et al.\cite{Liu2014a} Its operation is based on the valley splitting of the surface states on both sides of the (001) oriented thin film of a rock-salt TCI material. For a certain range of film thicknesses one dimensional gapless edge states appear protected by mirror plane symmetry of the film. These conduction channels can be shut by external electric field breaking the mirror symmetry and opening the gap in the edge states. If the valley splitting of surface states is reduced to zero the edge Dirac points would be located at a time reversal symmetry point. This protects the gap closing against electric field, thus rendering the device inoperative. Ref.~\onlinecite{Polley2018} demonstrates that the same argument can be made about roughness of the surface of a NI overlayer encapsulating the TCI film, or the NI/TCI interface.

\section{Conclusions}

The study presented here shows that the atomic steps on (001) surfaces of IV-VI TCIs greatly affect the topologically protected surface states. The proposed EF model relates both the observed collapse of the Dirac cone splitting on a rough surface and the formation of 1D zero-modes at the step edges, to the intervalley interaction. Our approach reproduces the results of exhaustive tight binding calculations\cite{Sessi2016,Polley2018} and confirms the insight of the original minimal toy model,\cite{Sessi2016} thus constituting a golden mean between the two. The solutions at isolated [11] and [01] step edges can be easily evaluated to serve as a footing for further study, e.g., of the electronic correlations in such a 1D system. A recent observation of a zero-bias peak of conductance in a low temperature measurement of the (001) and (011) surfaces of (Pb,Sn,Mn)Te with atomic steps motivates such theoretical research.\cite{Mazur2017}

Our article focuses on a specific surface of (Pb,Sn)Se, but the method itself is more general. Whereas the quantitatively accurate model is easily applicable to every TCIs of rock-salt structure, the strategy of combining realistic tight binding calculations with $\mathbf{k} \cdot \mathbf{p}$ symmetry based analysis can be implemented to any compound in which surface steps define electronic domains of distinct topology. In this approach the tight binding eigenstates are cast onto the basis of the continuous model, thus allowing quantification of the degrees of freedom relevant to the electronic connection between the terraces. This facilitates the determination of topological indices characterizing adjacent terraces and their interfaces, enabling recognition of possible 1D edge channels.

In our case the neighboring terraces are described by a different value of winding number, leading to 1D states on the step edges. In this way our result bears resemblance to second order topological insulators, where 1D hinges between two insulating surfaces host linearly dispersing topologically protected states.\cite{Schindler2018} (001) surfaces of IV-VI TCIs are also predicted to host 1D channels localized on walls between domains characterized by a different ferroelectric distorsion.\cite{Serbyn2014} However, those cases differ from the surface with step edges, where the adjacent electronic domains are intrinsically gapless. The states confined to the step edges in (Pb,Sn)Se should be also distinguished from similar modes recently observed in Bi$_2$Se$_3$,\cite{Fedotov2018} which are predicted to derive from a potential dip at the quintuple-layer steps.\cite{Xu2018}

\begin{acknowledgments}
We thank Per\l{}a Kacman, Jakub Tworzyd\l{}o and \L{}ukasz Cywi\'{n}ski for helpful discussions. This work was supported by the Polish National Science Centre under projects 2014/15/B/ST3/03833 and 2016/23/B/ST3/03725. Tight binding calculations were carried out at the Academic Computer Centre in Gda\'{n}sk.
\end{acknowledgments}

\end{document}


\title{Supplemental Material to 'Topological states on uneven (Pb,Sn)Se (001) surfaces'}
\author{Rafa\l{} Rechci\'nski}
\author{Ryszard Buczko}%
\affiliation{%
 Institute of Physics, Polish Academy of Sciences, Aleja Lotnikow 32/46, PL-02668 Warsaw, Poland
}%

\maketitle

\section{The formula for winding number}
The Hamiltonian $H$ can be characterized by the winding number $\nu$ topological invariant if it has the property
\begin{equation}
\Gamma H \Gamma  = -H,
\end{equation}
where $\Gamma = \Gamma^\dagger = \Gamma^{-1}$ is the chiral symmetry operator.\cite{Chiu2016} $\Gamma$ defines two chiral subspaces associated with projectors $P_+$ and $P_-$. The projectors are related to $\Gamma$ by the following formulas:
\begin{equation}
 \Gamma = P_+ - P_-, \quad I = P_+ + P_-, \quad P_+ = \frac{1}{2}\left( I + \Gamma\right), \quad P_- = \frac{1}{2}\left( I - \Gamma\right). 
 \end{equation}
It's easy to see that $H$ equals zero within those subspaces, but has matrix elements mixing "$+$" and "$-$".

We evaluate the winding number $\nu$ using the formula
\begin{equation} \label{eq:winding_formula}
\nu = - \frac{1}{2 \pi} \Im \int dk \frac{ \partial }{\partial k} \log \det h,
\end{equation}
where the matrix $h$ is extracted from $H$ as 
\begin{equation}
h = P_+ H P_-.
\end{equation}
A similar equation was used in Ref.~\onlinecite{Fulga2012}, for instance. We note that the definition~\eqref{eq:winding_formula} is equivalent to the canonical formula \cite{Chiu2016}
\begin{equation} \label{eq:can_winding_formula}
\nu = \frac{i}{2 \pi} \int dk \, \Tr \left( q^{-1} \frac{ \partial q }{\partial k} \right),
\end{equation}
where $q$ is defined as $P_+ Q P_-$ and
\begin{equation}
Q =\sum_{n \in \textrm{cond.}} \left( \ket{n}\bra{n} - \Gamma \ket{n}\bra{n} \Gamma\right)
\end{equation}
is the Hamiltonian deformed adiabatically into the flat band limit (the sum runs over the states in the conduction band).

Below is the proof of equivalence of Eqs.~\eqref{eq:winding_formula} and~\eqref{eq:can_winding_formula}.

From the properties of trace we get
\begin{equation} \label{eq:tr_to_det}
\frac{ \partial }{\partial k} \log \det h = \frac{ \partial }{\partial k}  \Tr   \log h  =  \Tr \left( h^{-1} \frac{ \partial h }{\partial k} \right) . 
\end{equation}
Please note that in the basis of eigenstates of the operator $\Gamma$, the matrix $h$ constitutes an off diagonal part of the Hamiltonian. Its inverse is equal to:
\begin{equation} \label{eq:analog}
h^{-1} = P_- H^{-1} P_+.
\end{equation}
Therefore, we have
\begin{equation} \label{eq:tr_with_P}
\Tr \left( h^{-1} \frac{ \partial h }{\partial k} \right) = \Tr \left( P_- H^{-1} P_+ \frac{ \partial H }{\partial k} P_- \right) .
\end{equation}
$H$ in its eigenbasis has the form
\begin{equation}
H= \sum_{n \in \textrm{cond.}} E_n \left( \ket{n}\bra{n} - \Gamma \ket{n}\bra{n} \Gamma\right).
\end{equation}
The operators in the trace in Eq.~\eqref{eq:tr_with_P} can be easily evaluated:
\begin{equation}
P_- H^{-1} P_+ = 2 \sum_{n \in \textrm{cond.}} E^{-1}_n \left( P_- \ket{n}\bra{n} P_+ \right),
\end{equation}
\begin{equation}
P_+ \frac{ \partial H }{\partial k} P_- = 2 \sum_{n \in \textrm{cond.}} 
\frac{ \partial E_n }{\partial k}  P_+ \ket{n}\bra{n} P_-  + 2 \sum_{n \in \textrm{cond.}} E_n \left( P_+ \frac{ \partial \ket{n} }{\partial k}\bra{n} P_- + P_+ \ket{n}\frac{ \partial \bra{n} }{\partial k} P_-\right) .
\end{equation}

Their product is
\begin{multline}
P_- H^{-1} P_+ \frac{ \partial H }{\partial k} P_- = \\
2 \sum_{n \in \textrm{cond.}} \left[  E^{-1}_n \frac{ \partial E_n }{\partial k} \left( P_- \ket{n}\bra{n} P_- \right) +   P_- \ket{n}\frac{ \partial \bra{n} }{\partial k} P_- + 2 \sum_{m \in \textrm{cond.}}  E^{-1}_n E_m \left( P_- \ket{n}\bra{n} P_+  \frac{ \partial \ket{m} }{\partial k}\bra{m} P_- \right)  \right ] ,
\end{multline}
which can be calculated using the fact that for $\ket{n}$ and $\ket{m}$ in the conduction band the matrix element equals:
\begin{equation}
\bra{n} P_{\pm} \ket{m} = \frac{1}{2} \Braket{n|m} \pm \frac{1}{2} \Braket{n|\Gamma|m} = \frac{1}{2}\delta_{nm}
\end{equation}
(the second term is zero, because if $\ket{n}$ and $\ket{m}$ are in the conduction band then $\Gamma \ket{m}$ is in the valence band and therefore orthogonal to $\ket{n}$).

Taking the trace we find
\begin{equation} \label{eq:result}
\Tr \left( h^{-1} \frac{ \partial h }{\partial k} \right) = \sum_{n \in \textrm{cond.}} \left[ \frac{ \partial \log E_n }{\partial k} + 2i \Im \bra{n} \Gamma  \frac{ \partial \ket{n} }{\partial k} \right].
\end{equation}

If we now repeat the evaluation of $\Tr \left( h^{-1} \frac{ \partial h }{\partial k} \right)$ but plugging $Q$ instead of $H$ we can rewrite the result \eqref{eq:result} as
\begin{equation}
\Tr \left( q^{-1} \frac{ \partial q }{\partial k} \right) = 2 i \Im \sum_{n \in \textrm{cond.}} \bra{n} \Gamma  \frac{ \partial \ket{n} }{\partial k} = i \Im \Tr \left( h^{-1} \frac{ \partial h }{\partial k} \right),
\end{equation}
since in the flat band limit energies $E_n$ are constant across the $k$ space. This concludes the proof of the equivalence of Eqs.~\eqref{eq:winding_formula} and~\eqref{eq:can_winding_formula}.

\section{Winding number in the envelope function model}

\begin{figure}[h]
\centering
\includegraphics[width=0.6\textwidth]{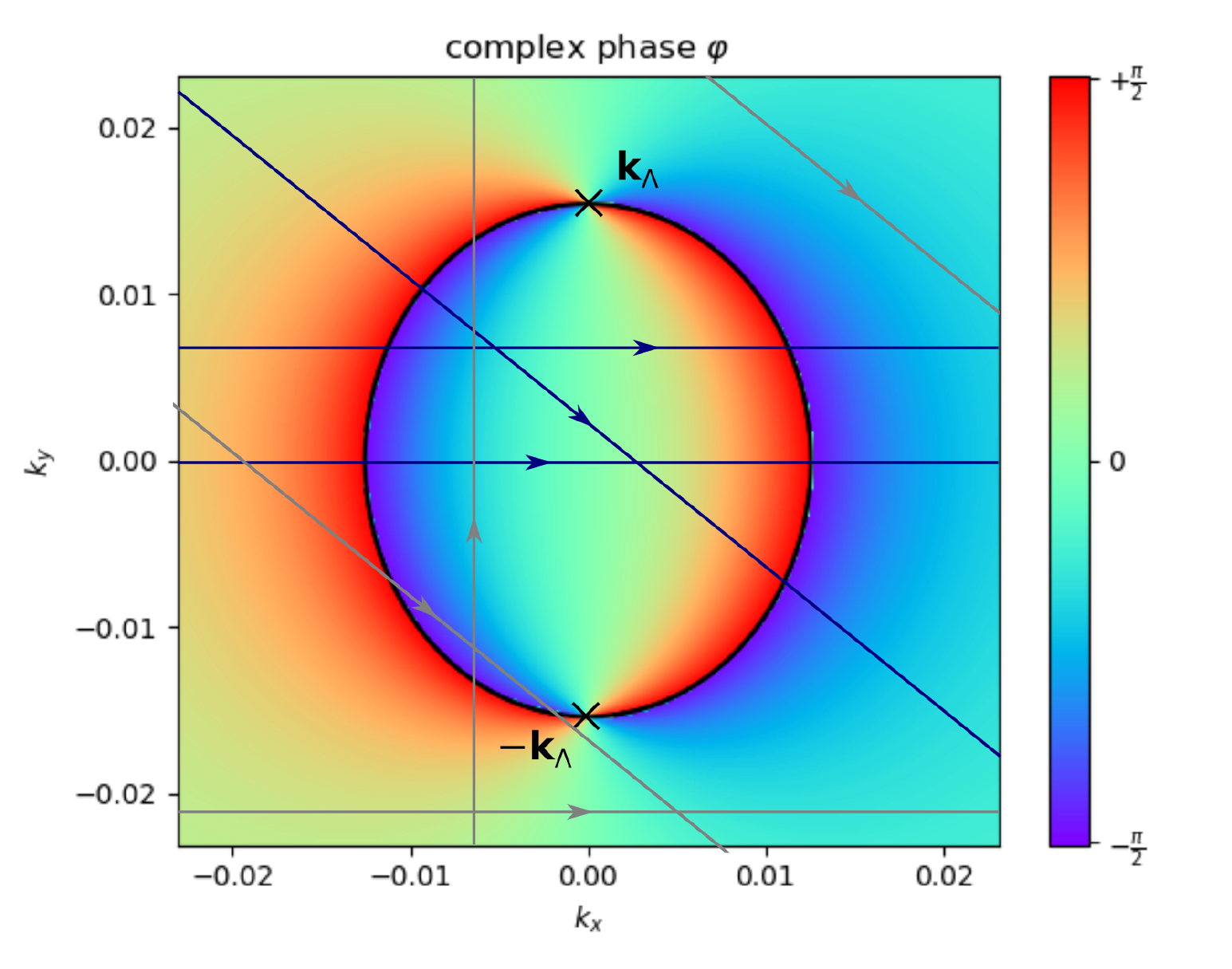}
\caption{Plot of function~\eqref{eq:phase} for $m = 0.036, v_y = 2.3, v_x = 2.5, v''_x = 1.4$, and $\eta = 0$. The top and bottom quadrant points of the ellipse correspond to Dirac points. The blue lines have winding number $\nu = -1$, while the grey ones are topologically trivial. $[11]$ steps correspond to vertical and horizontal lines (describing vicinities of both $\overline{X}$ points). $[10]$ steps correspond to diagonal lines.} \label{fig:phase}
\end{figure}

To describe the surface states we use a simple envelope function model with the chiral symmetric $\mathbf{ k} \cdot \mathbf{ p}$ Hamiltonian
\begin{equation} \label{eq:kpHamiltonian}
H( \mathbf{ k} ) = m \tau_x + k_x (v_x s_y + v''_x \tau_z s_z) - k_y v_y s_x,
\end{equation}
where $\mathbf{ k}=0$ is the $\Xbar$ point. We can pick the chiral symmetry operator $\Gamma = \tau_y s_z$ such that $\Gamma H \Gamma  = -H$. By applying a rotation $U = \frac{1}{\sqrt{2}} (\tau_z + \tau_y )$ we switch to a basis in which $\Gamma$ is diagonal $\Gamma' = \tau_z s_z$, and the Hamiltonian takes the form
\begin{equation}
H'(k_x,k_y) = - m \tau_x + k_x (v_x s_y + v''_x \tau_y s_z) - k_y v_y s_x .
\end{equation}
The operator $\Gamma'$ defines two chiral subspaces:
\begin{itemize}
\item "$+$" spanned by $\ket{\tau=1, s=1}, \ket{\tau=-1, s=-1} $,
\item "$-$" spanned by $\ket{\tau=1, s=-1}, \ket{\tau=-1, s=1} $.
\end{itemize}
$H'$ equals zero within those subspaces, but has matrix elements mixing "$+$" and "$-$". We can evaluate
\begin{itemize}
\item $\Braket{1,1| H' |1,-1} = -i k_x v_x -k_y v_y$,
\item $\Braket{1,1| H' |-1,1} = -m -i v''_x k_x$,
\item $\Braket{-1,-1| H' |1,-1} = -m - i v''_x k_x$,
\item $\Braket{-1,-1| H' |-1,1} = i k_x v_x - k_y v_y$,
\end{itemize}
and put it in  matrix form
\begin{equation}
h(k_x,k_y) = \begin{pmatrix}
-i k_x v_x -k_y v_y &  -m -i v''_x k_x \\
-m - i v''_x k_x & i k_x v_x - k_y v_y \\
\end{pmatrix} = 
-k_y v_y \sigma_0 - i k_x v_x \sigma_z - (m +i v''_x k_x) \sigma_x.
\end{equation}
To calculate the winding number we will need the following expression
\begin{equation}
 \log \det h =  \log [ k^2_y v^2_y + k^2_x v^2_x - (m +i v''_x k_x)^2 ] = \mu + i \varphi,
\end{equation}
where
\begin{equation}
\mu = \log| k^2_y v^2_y + k^2_x v^2_x - (m +i v''_x k_x)^2  |
\end{equation}
and
\begin{equation} \label{eq:phase}
\varphi = \arctan\left( \frac{- 2 m v''_x k_x}{k^2_y v^2_y + k^2_x v^2_x -  m^2 + v''^2_x k^2_x}  \right) + \eta \pi , \quad (\eta = 0 \textrm{ or } \eta = 1).
\end{equation}
 The phase $\varphi$ has a branch cut on the ellipse defined by 
\begin{equation}
k^2_y v^2_y + k^2_x (v^2_x  + v''^2_x) =  m^2.
\end{equation}
For large $|\mathbf{ k}|$ the value of $\varphi$ approaches $\eta \pi$. Plot of the function $\varphi$ (for $m,v''_x>0$, and $\eta = 0$) is shown on Fig.~\ref{fig:phase}. 

Let us consider a straight line defined by $k_y = \mathrm{ const.}$ directed towards positive infinity. If the line does not enter the area between the two Dirac points, $\varphi$ is smooth and eventually acquires the same value at $k_x = \pm \infty$. If the line crosses between the Dirac points (through the ellipse) it changes smoothly by $2 \pi$. According to formula~\eqref{eq:winding_formula} this corresponds to winding number $\nu = -1$. Note that in Fig.~\ref{fig:phase} the smooth increase of $\varphi$ is interrupted by two $-\pi$ jumps. This is due to the plot showing just the $\eta=0$ branch. Such a jump can be represented as a smooth change of $\varphi$ between the $\eta = 0$ and $\eta = 1$ branches of the solution~\eqref{eq:phase}.

The function $\varphi$ is odd with respect to $m$, which means that for a system with opposite sign of $m$ the phases are $-2 \pi$ (winding number $\nu = 1$) between the Dirac points and $0$ outside of them.

Although the strict definition of $\nu$ requires that it's calculated along a closed contour, e.g., a 1D cut through a Brillouin zone, we believe that in our case integration along $k_x \in (-\infty,\infty)$ gives valid results. For $k_x$ away from $\Xbar$ the surface states are hybrydized into bulk bands, which are well separated in energy. No nontrivial topology of surface states is expected outside the vicinity of $\Xbar$ points.

In conclusion, the $m$ and $-m$ regions are described by winding numbers $-1$ and $+1$. An interface between those two media can host two topologically protected states. Due to chiral symmetry their energies are necessarily $E$ and $-E$. In the next section we will show, that in our system $E=0$.

\section{Zero modes}
The envelope function equation of an atomic step can be written as
\begin{equation}
(m \sign{x_\perp}) \tau_x \Psi  + \chi_\parallel k_\parallel \Psi -i \chi_\perp \frac{d \Psi }{d x_\perp} = E \Psi,
\end{equation}
where $x_\perp$ is the axis perpendicular to the atomic step, and $\chi_\perp$ and $\chi_\parallel$ are defined by an appropriate rotation of $k_x (v_x s_y + v''_x \tau_y s_z) - k_y v_y s_x$. We can write now
\begin{equation}~\label{eq:efe}
 -i  \chi_\perp \frac{d \Psi }{d x_\perp} =   (E - m \sign{x_\perp} \tau_x - \chi_\parallel k_\parallel) \Psi.
\end{equation}

Let us act on Eq.~\eqref{eq:efe} with a projection operator associated with the "$+$" subspace.
\begin{equation}
 -i P_+  \chi_\perp P_- \frac{d P_- \Psi }{d x_\perp} + P_+ (m \sign{x_\perp} \tau_x + \chi_\parallel k_\parallel) P_- \Psi =   E P_+ \Psi .
\end{equation}
If this equation has a solution for $E=0$, then it exists only in the "$-$" subspace.
\begin{equation} \label{eq:envEqForPB}
 -i P_+  \chi_\perp P_- \frac{d P_- \Psi }{d x_\perp} + P_+ (m \sign{x_\perp} \tau_x + \chi_\parallel k_\parallel) P_- \Psi = 0  .
\end{equation}
We will denote
\begin{equation}
h_\perp = P_+  \chi_\perp P_-, \quad h_\parallel = P_+ \chi_\parallel P_-, \quad h_0(x_\perp) = P_+ (m \sign{x_\perp}) \tau_x P_-, \quad \psi = P_- \Psi.
\end{equation}
In our case, explicitly
\begin{equation}
h_0(x_\perp) = - (m \sign{x_\perp}) \sigma_x, \quad h_x = - i v_x \sigma_z - i v''_x  \sigma_x, \quad h_y = - v_y \sigma_0,
\end{equation}
\begin{equation}
h_\perp = \cos \theta h_x + \sin \theta h_y, \quad h_\parallel = -\sin \theta h_x + \cos \theta h_y,
\end{equation}
where $\theta$ is the angle between $[11]$ and the direction of the step edge.

The equation becomes
\begin{equation}
 -i h_\perp \frac{d \psi }{d x_\perp} + h_0(x_\perp) \psi + h_\parallel k_\parallel \psi = 0  .
\end{equation}
We look for independent modes if $x_\perp>0$ (denoted below by a subscript "$>$")
\begin{equation}
\psi(x_\perp) = e^{ \lambda_> x_\perp  } f_>,
\end{equation}
where $f_>$ is $\psi$ at $x_\perp=0$. The modes are solutions of the eigenequation:
\begin{equation} \label{eq:eqForMgr}
  - i h^{-1}_\perp \left[- m \sigma_x + h_\parallel k_\parallel \right] f_>  = \lambda_> f_>  .
\end{equation}

The modes for $x_\perp<0$ can be found by solving the equation after substituting $m \rightarrow -m$ or $h_0^> \rightarrow h_0^< = -h_0^>$. The solutions can be obtained from the solutions of~\eqref{eq:eqForMgr}, if we notice that
\begin{equation}
   - i  h^{-1}_\perp  \left[ m \sigma_x  +  h_\parallel  k_\parallel \right]   \sigma_y f^*_>  =   - \lambda_>^* \sigma_y f^*_>.
\end{equation}
Thus we get
\begin{equation} \label{eq:XMniejszeZWiekszego}
\lambda_< = - \lambda_>^* , \quad   f_< =  \sigma_y f^*_>.
\end{equation}

Let us assume two eigenvalues $\lambda_>$ exist and they both have a negative real part (describe evanescent modes on $x_\perp>0$ half-plane). Then $\lambda_<$ will have positive real parts, which will describe modes evanescent on $x_\perp<0$. The eigenvectors of any of these matrices, if linearly independent, will span the entire $\mathbb{C}^2$ space, which means that the continuity condition between modes at both half-planes will always be satisfied. In such a case there will be two zero modes localized at the step edge and both will belong to the same chiral subspace "$-$".

According to the winding number argument, there are at most two localized modes at the step edge. We can therefore conclude that the subspace "$+$" doesn't host any interface states. To verify this we investigate an equation analogous to Eq.~\eqref{eq:envEqForPB} for the "$+$" subspace
\begin{equation} \label{eq:envEqForPA}
 -i P_-  \chi_\perp P_+ \frac{d P_+ \Psi }{d x_\perp} + P_- (m \sign{x_\perp} \tau_x + \chi_\parallel k_\parallel) P_+ \Psi = 0  .
\end{equation}
The equation for a zero-energy mode $\psi' = P_+ \Psi$ is
\begin{equation}
 -i h^\dagger_\perp \frac{d \psi'}{d x_\perp} + h^\dagger_0(x_\perp) \psi' + h^\dagger_\parallel k_\parallel \psi' = 0  ,
\end{equation}
where
\begin{equation}
h^\dagger_0(x_\perp) = h_0^*(x_\perp) =  h_0(x_\perp), \quad h^\dagger_x = h^*_x = -h_x, \quad h^\dagger_y = h^*_y = h_y.
\end{equation}
The equation for eigenmodes can be now obtained by transforming~\eqref{eq:eqForMgr} to get
\begin{equation}
 -  i (h^{-1}_\perp)^* \left[ -m \sigma_x + h^*_\parallel k_\parallel \right] f^*_>   = - \lambda_>^* f^*_>.
\end{equation}
This shows that
\begin{equation}
\lambda'_> = - \lambda_>^*, \quad f'_>  =  f^*_> ,
\end{equation}
where the primed quantities belong to the "$+$" subspace, while the unprimed ones to the "$-$" subspace. We can see now that if $\lambda_>$ describe evanescent modes then $\lambda'_>$ modes explode in the $x_\perp\rightarrow \infty$ limit and cannot form a localized state. Thus, zero-energy modes exist only in the "$-$" subspace.

Let us note that the designation of "$+$" and "$-$" is arbitrary, as their role changes upon changing the gauge: $m(x_\perp) \rightarrow - m(x_\perp)$ or $\Gamma \rightarrow - \Gamma$.

We will now check the simplest case $h_\perp = h_x$, $h_\parallel = h_y$, $k_y=0$. The eigenequation in the "$-$" subspace for $x>0$ is
\begin{equation}
 - \frac{ m }{ (v_x^2 + v''^2_x)  } (v''_x \sigma_0 + i v_x \sigma_y ) f_> = \lambda_> f_>  .
\end{equation}
The eigenvalues and eigenvectors are
\begin{equation}
\lambda^\pm_> = \frac{ -m v''_x  \pm i m v_x }{  (v_x^2 + v''^2_x) }, \quad
f^\pm_> = \begin{pmatrix}
\pm i \\ 1
\end{pmatrix}.
\end{equation}

Without loss of generality we set $m>0$. Then for both eigenvalues $\Re{\lambda}<0$, which means that in this case the two zero modes localized at the step edge belong to the chiral subspace "$-$". Let us recall now that states limited to one chiral subspace necessarily have zero energy. 

Solutions for $x<0$ are
\begin{equation}
\lambda^\pm_< = \frac{ +m v''_x  \mp i m v_x }{  (v_x^2 + v''^2_x) }, \quad
f^\pm_< = \begin{pmatrix}
\pm i \\ 1
\end{pmatrix}.
\end{equation}
The continuity condition is satisfied automatically as $f^\pm_< = f^\pm_>$. We can write the two solutions explicitly
\begin{equation}
\psi^\pm(x) = \sqrt{ \frac{m v''_x}{ 2 (v_x^2 + v''^2_x)  } }   e^{ \frac{ -m v''_x  \pm i m v_x }{  (v_x^2 + v''^2_x) } |x| } \begin{pmatrix}
\pm i \\ 1
\end{pmatrix}.
\end{equation}

The case of $k_y \neq 0$ is more difficult to solve explicitly. Still, the existence of zero modes can be proven by considering the addition of a $k_y$-dependent term to the Hamiltonian as an adiabiatic deformation. Since the term doesn't break chiral symmetry, both zero-modes will still belong to subspace "$-$" and will not change their energy, provided that $|k_y|< \frac{m}{v_y} = |\mathbf{ k}_\Lambda|$. For larger $|k_y|$ the winding number is 0 (see Fig.~\ref{fig:phase}) and the localized states vanish completely. Note that if we go through one of the $|k_y|= \frac{m}{v_y}$ points we encounter closing of the band-gap and the deformation is no longer adiabatic.

The very same approach can be used to show that the zero modes exist for different directions of the step edge, as rotation of the Hamiltonian in $k$ space can also be understood as a slow deformation. 

All such deformations can be seen as transformations to straight lines drawn on the complex phase plot (Fig.~\ref{fig:phase}) from which we conclude that the zero-energy modes exist if the corresponding straight line in $k$ space passes between the two Dirac points at $\pm \mathbf{ k}_\Lambda$, i.e., between projections of Dirac points onto the step edge Brillouin zone.

The case of $[10]$ direction of the step edge requires more attention. If we consider a straight line which passes through $\Xbar_1$, rotated from direction $[11]$ to $[01]$ we encounter no closing of the gap of the $\kp$ Hamiltonian~\eqref{eq:kpHamiltonian}. However, in a description involving the entire surface Brillouin zone, such a rotating straight line would enter the vicinity of $\Xbar_2$, which features gap closings analogous to the vicinity $\Xbar_1$. Nontheless, our tight binding calculation for $[10]$ steps presented in the main text shows that there is no scattering between the $\Xbar_1$ and $\Xbar_2$ on the step edge. Therefore, we conclude that the vicinities of the two $\Xbar$ points can be treated in this calculation as seperate systems, each characterized by its own winding number.

\section{Correspondence between $\kp$ and tight binding models }
For realistic predictions of the electronic states at the (001) surface we use a nearest-neighbour 18-orbital sp$^3$d$^5$ tight binding (TB) model with spin-orbit interactions included. The solid solution (Pb,Sn)Se is modelled within the virtual crystal approximation (VCA) with parameters derived according to the procedure described in Ref.~\onlinecite{Wojek2013}, which includes effects of temperature and strain on the band-structure.  We choose parameters describing \ce{Pb_{0.68}Sn_{0.32}Se} at a temperature of $100\,\mathrm{K}$.

In this section we show how to establish one to one correspondence between the basis states of the $\kp$ Hamiltonian~\eqref{eq:kpHamiltonian}, and the (001) surface states of (Pb,Sn)Se expressed in the tight binding approximation. Such calculation is necessary in order to inspect how the basis states of~\eqref{eq:kpHamiltonian} behave in presence of atomic steps.

For the analysis we choose the point $\Xbar_1 = \pi/a_0 (1,1,0)$. The basis of~\eqref{eq:kpHamiltonian}, chosen after Ref.~\onlinecite{Liu2013}, is defined by eigenspaces of diagonal operators $\tau_z$ and $s_z$. $\tau_z$ distinguishes between surface states at $\Xbar_1$ that can be unambiguously assigned to either the ${\rm L}_1 = \pi/a_0 (1,1,1)$ or the $\mathrm{ L}_2 = \pi/a_0 (1,1,-1)$ valley of the 3D reciprocal space. Eigenstates of $s_z$ are Kramers partners within each of the valleys. Operators of $(1 \bar{1} 0)$ and $(110)$ mirror reflections ($M_x$ and $M_y$) and time reversal $\Theta$ are expressed as
\begin{equation} \label{eq:symm_ops}
M_x = -i s_x, \quad M_y = -i \tau_x s_y, \quad \Theta = i s_y K.
\end{equation}

We begin our procedure by finding the eigenstates $\ket{ \Phi_i ( \mathbf{k}_{\Xbar_1}) }$ of the TB Hamiltonian of a thick slab, which are the four topological states on a selected surface, at the point $\Xbar_1$. Next we create linear combinations of $\ket{ \Phi_i (\mathbf{k}_{\Xbar_1}) }$ that can be assigned as eigenstates of $\tau_z$ and $s_z$.

The lattice of a semi-infinite crystal is spanned by the primitive vectors $\mathbf{ T} = l \mathbf{ t}_1 + m \mathbf{ t}_2 + n \mathbf{ t}_3$ with integer $l,m,n$, with a restriction that $n \geq 0 $. We choose $\mathbf{ t}_1 = \frac{a_0}{2} (1,-1,0)$, $\mathbf{ t}_2 = \frac{a_0}{2} (1,1,0)$, and $\mathbf{ t}_3 = \frac{a_0}{2} (1,0,1)$, which results in $(001)$ termination. In the tight binding approximation, ignoring the normalizing factor, we can write the surface state as
\begin{equation}
\Phi(\mathbf{ k}_{\bar X},\mathbf{ r}) = \sum_\alpha \sum_{n=0}^{\infty} a^{\alpha,n}_{\mathbf{ k}_{\bar X}} \sum_{\mathbf{ t}_{\parallel}}  e^{i \mathbf{ k}_{\bar X} \cdot ( \mathbf{ t}_{\parallel} + n \mathbf{ t}_3 + \mathbf{ d}_\alpha  ) } \phi_\alpha (\mathbf{ r} - \mathbf{ t}_{\parallel} - n \mathbf{ t}_3 - \mathbf{ d}_\alpha ),
\end{equation}
where $\mathbf{ t}_{\parallel}$ are vectors $l \mathbf{ t}_1 + m \mathbf{ t}_2$, $\phi_\alpha$ are the orthonormal L\"owdin orbitals centered at each atom, $a^{\alpha,n}_{\mathbf{ k}_{\bar X}}$ are the calculated tight binding coefficients ($\alpha$ runs through spin-orbitals on both cation and anion in a single layer, $n$ counts the layers). $\mathbf{ d}_\alpha$ equals zero for a cation and $\frac{a_0}{2} (1,0,0)$ for an anion.

To find the contributions of the $\mathrm{L}$ valleys into states $\Phi$ we need to express them in reciprocal space representation, which is achieved by finding overlaps of $\Phi$ with 3D Bloch waves describing various quasimomenta $\mathbf{k}$. The unnormalized 3D Bloch sum for a spin-orbital $\alpha$ is given by
\begin{equation}
\psi_\alpha (\mathbf{ k},\mathbf{ r}) = \sum_{\mathbf{ t}_{\parallel}} \sum_{n=-\infty}^{\infty} e^{i \mathbf{ k} \cdot ( \mathbf{ t}_{\parallel} + n \mathbf{ t}_3 + \mathbf{ d}_\alpha  ) } \phi_\alpha (\mathbf{ r} - \mathbf{ t}_{\parallel} - n \mathbf{ t}_3 - \mathbf{ d}_\alpha ).
\end{equation}
The overlap coefficient amounts to a Fourier transform of the tight binding amplitudes
\begin{equation} \label{eq:fourier}
\int d^3 r \, \psi^*_\alpha (\mathbf{ k},\mathbf{ r}) \Phi(\mathbf{ k}_{\bar X},\mathbf{ r}) \propto 
  \delta(\mathbf{ k}_\parallel - \mathbf{ k}_{\bar X}) \sum_{n=0}^\infty a^{\alpha,n}_{\mathbf{ k}_{\bar X}} e^{-i n \frac{a_0}{2}  k_\perp },
\end{equation}
with $\mathbf{ k} = \mathbf{ k}_\parallel  + \mathbf{ k}_\perp$, where $\mathbf{ k}_\parallel$ is lying on the plane of the surface Brillouin zone, and $\mathbf{ k}_\perp$ is perpendicular to it. The surface state $\Phi (\mathbf{k}_{\Xbar_1}, \mathbf{r})$ is constructed only of Bloch waves corresponding to quasimomenta $\mathbf{k} = (\pi/a_0,\pi/a_0,k_\perp)$.

\begin{figure}
\includegraphics[scale=1]{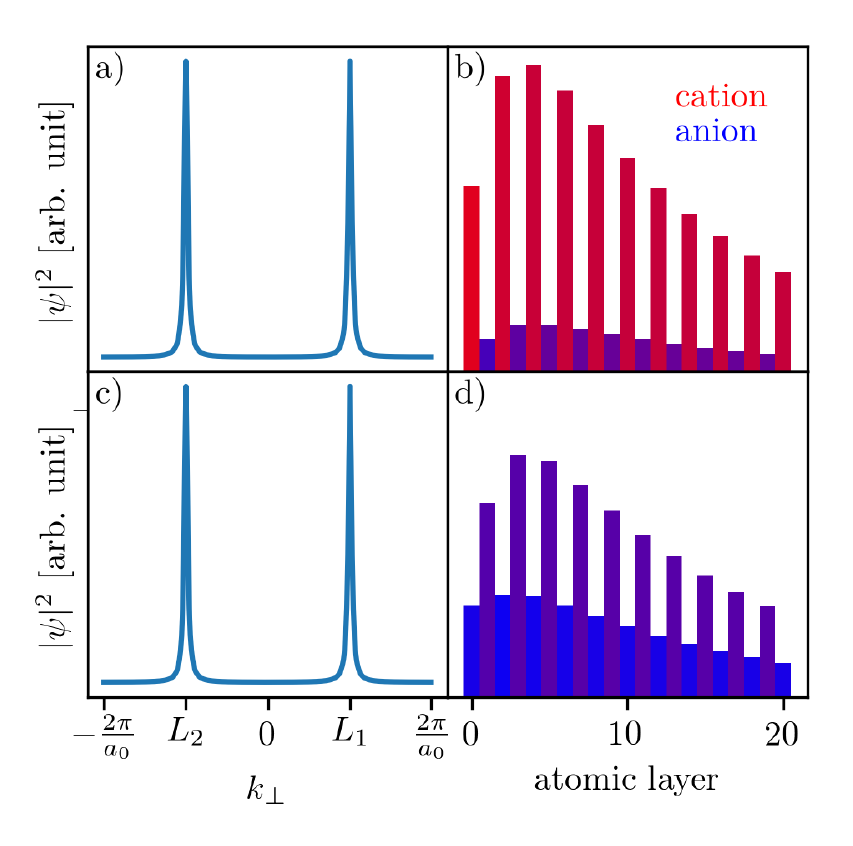}
\caption{The tight binding wavefunctions of the valley-split (001) surface states of Pb$_{0.68}$Sn$_{0.68}$Se at $T=100\,\mathrm{ K}$ calculated at the $\Xbar$ point of the flat surface. Panels (a) and (c) show squared moduli of Fourier transforms along the axis perpendicular to the surface, while (b) and (d) depict squared moduli of wavefunction amplitudes at the first 21 layers near the surface. The bonding state (a, b) is composed mostly of cationic $p$ orbitals on even layers (with the surface layer labelled 0) and anionic $p_z$ orbitals on odd layers. The antibonding state (c, d) is made of anionic $p_x$ and $p_y$, and cationic $s$ orbitals on odd layers and anionic $p_z$ on even layers. The orbital make-up of the states at individual layers is denoted by color of the bar ranging from red for cations only to blue for anions only. Only two out of four states are presented, as their Kramers partners have identical spatial probability distributions. } \label{fig:states_split}
\end{figure}

The formula~\eqref{eq:fourier} can be easily computed numerically using the discrete Fourier transform, as the surface states vanish exponentially in the bulk of the crystal, and the coefficients $a^{\alpha,n}_{\mathbf{ k}_{\bar X}}$ can be cut off after a certain large $n$. Figure~\ref{fig:states_split}(a,c) depicts results of such an analysis for surface states of Pb$_{0.68}$Sn$_{0.68}$Se, which shows that the states originate from the vicinities of the two L points. In real space (Fig.~\ref{fig:states_split}(b,d)) the two states have profiles resembling interference fringes with a spacing of one lattice constant $a_0$ (two monolayers). This is consistent with a superposition of two waves that are separated by a wavevector of length $2 \pi / a_0$ in the direction perpendicular to the surface.

To find superpositions of $ \ket{ \Phi_i (\mathbf{ k}_{\Xbar_1}) }$ belonging to each of the valleys we introduce a valley separating operator $ \Xi = P_> - P_< $, where $P_>$ is the projection onto the subspace of $k_\perp  \in [0,2 \pi/a_0]$, and $P_< = 1 - P_>$, and compute the matrix elements $\braket{\Phi_i (\mathbf{ k}_{\Xbar_1}) | \Xi | \Phi_j (\mathbf{ k}_{\Xbar_1}) }$. The states diagonalizing the resulting matrix are primarily composed of states from one valley (see Fig.~3 in the main text of the paper). Those valley-resolved states we assign as eigenvectors of $\tau_z$ (and basis vectors of the $\tau$ subspace). Within the eigenspaces of $\tau_z$ we find $\pm 1$ eigenstates of $s_z$. We use the fact, that the basis states should be related by symmetry operations defined in~\eqref{eq:symm_ops}. Therefore we express the same operations in the tight binding basis and analyze their action on the valley-resolved states. Then we find such linear combinations of valley-resolved states that transform according to~\eqref{eq:symm_ops}. 

In this way we get TB states $\{ \ket{ F^s_{\mathrm{ L}_i}  } \} _{i\in\{1,2 \}, s = \pm 1} $ that have the properties of basis states of Hamiltonian~\eqref{eq:kpHamiltonian}. Their probability densities are shown in Fig.~3 in the main text of the paper.

Now we can recreate the $\kp$ Hamiltonian from the TB results. To that end we perform additional TB calculations of surface states $\ket{ \Phi_i (\mathbf{ k}) }$ for several points in the vicinity of $\Xbar_1$. At each of the points the Hamiltonian restricted to the subspace of surface states can be written in its eigenbasis as 
\begin{equation} \label{eq:Hkdiag}
H^\mathrm{ diag}(\mathbf{ k}) = \sum_{i=1}^4 \epsilon_i (\mathbf{ k}) \ket{ \Phi_i (\mathbf{ k}) } \bra{ \Phi_i (\mathbf{ k}) },
\end{equation}
where $\epsilon_i (\mathbf{ k})$ and $e^{i \mathbf{ k} \cdot \mathbf{ r}} \ket{ \Phi_i (\mathbf{ k}) }$ are the computed eigenvalues and eigenvectors respectively. The matrices~\eqref{eq:Hkdiag} can be now expressed in the new basis $\ket{ F^s_{\mathrm{ L}_i}  }$. For each of the $\mathbf{ k}$ points we first confirm that the completeness relation $\sum_{i,s} |\braket{\Phi_j (\mathbf{ k})| F^s_{\mathrm{ L}_i}  }|^2 = 1 $ holds for every $j$. 

We end up with a set of matrices $H(\mathbf{ k})$ expressed in the basis defined at $\Xbar_1$, which can be written, up to first order in $\bf k$, as~\eqref{eq:kpHamiltonian}. While our analysis of the vicinity of $\Xbar$ shows that the term $v''_y \tau_y k_y$ can be non-zero as well, we set it to zero for a better agreement of the dispersion to the tight binding result for large $\mathbf{ k}$. All other terms allowed by symmetry, listed in the Supplemental Material of Ref.~\onlinecite{Liu2013}, turn out to equal zero in our numerical result.

Note that due to gauge freedom in defining $H(\mathbf{k})$ our procedure does not ensure that~\eqref{eq:Hkdiag} expressed in the basis  $\ket{ F^s_{\mathrm{ L}_i}  }$ will have the form~\eqref{eq:kpHamiltonian}. However, the desired form can be retrieved by gauge transformation $U_\theta=\exp( i \theta \tau_z s_x )$ with a properly chosen angle $\theta$.

Thus we have estabilished correspondence between the $\kp$ model and the TB solutions for the (Pb,Sn)Se (001) surface states.

\section{Comparison of the envelope function model with tight binding results}
\begin{figure}
\centering
\includegraphics[width=\textwidth]{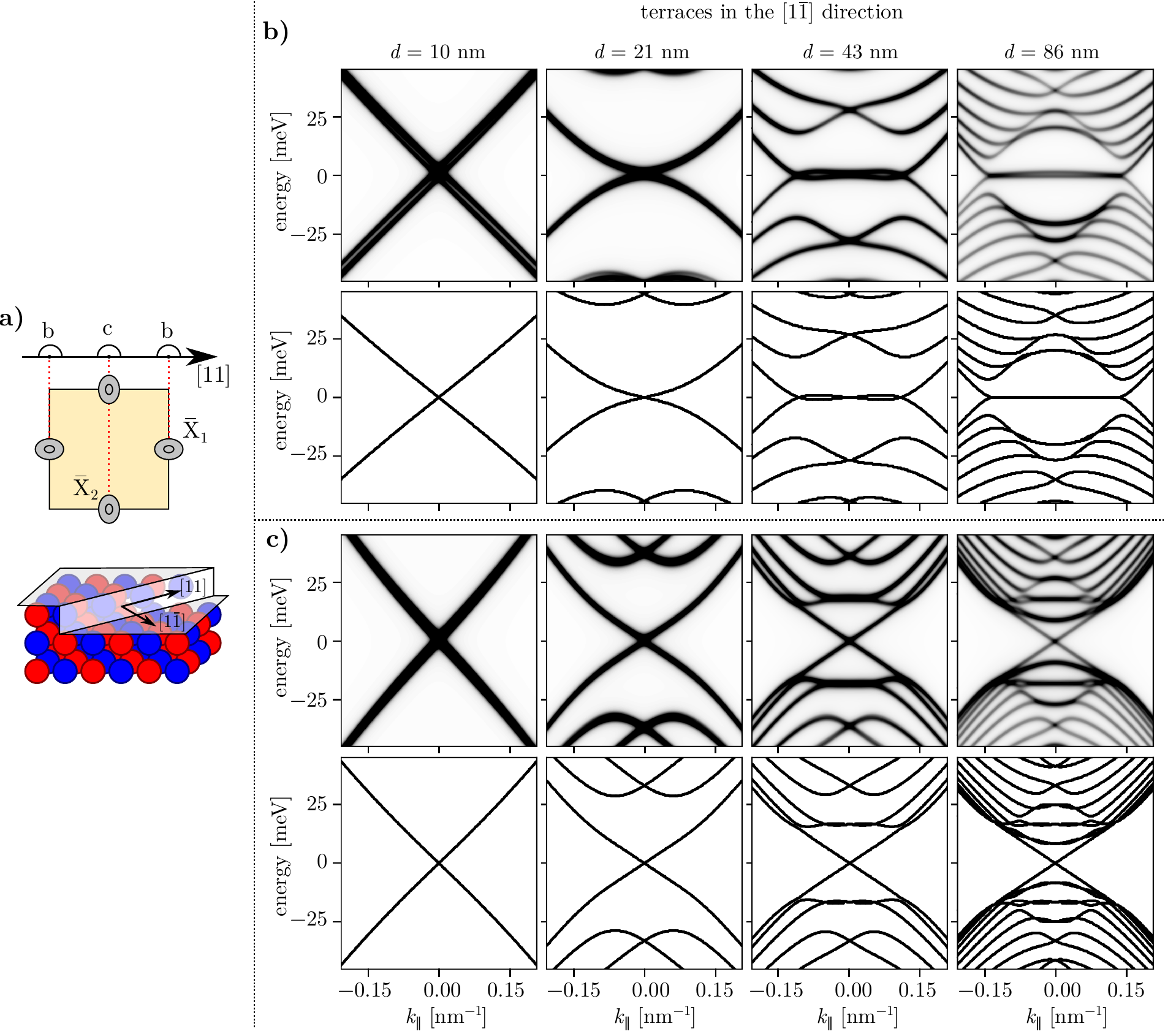}
\caption{Electronic spectrum of a (001) \ce{Pb_{0.68}Sn_{0.32}Se} surface in the presence of evenly distributed, equally wide, parallel terraces of one monolayer height. Above each plot $d=d_1=d_2$ defines the step-to-step distances. a) Alignment of [11] steps with respect to the surface Brillouin zone and the crystal structure. b), c) Spectra obtained for the vicinity of $\Xbar_1$ (b) and $\Xbar_2$ (c) for periodic arrays of [11] steps with different $d$. In (b) and (c) the top rows are tight binding results, and bottom rows are results obtained within the envelope function approximation. A similar comparison for [10] steps is shown in the main text.} \label{fig:comp}
\end{figure}

In this section we show spectra of (001) surface states in the presence of evenly distributed, equally wide, parallel terraces of one monolayer height. In Fig.~\ref{fig:comp} results of the simple envelope function approximation are compared to surface spectral densities obtained from a realistic tight binding model. All results are for \ce{Pb_{0.68}Sn_{0.32}Se} at temperature $100\,\mathrm{ K}$, which is described by parameters $m=36\,\mathrm{ meV}$, $v_x = 0.24\,\mathrm{ eV \cdot nm}$, $v''_x =0.13 \,\mathrm{ eV \cdot nm}$, $v_y = 0.22 \,\mathrm{ eV \cdot nm}$. Tight binding parameters are chosen according to the procedure described in Ref.~\onlinecite{Wojek2013}.

The surface spectral function, a result analogous to experimental ARPES spectra, is obtained for flat surfaces of semi-infinite crystals by means of the recursive Green's function method described in Ref.~\onlinecite{LopezSancho1985}. The Green's function of the terraces $G_\mathrm{ terrace }(E,\mathbf{ k})$ can be calculated by plugging the flat-surface Green's function $G_\mathrm{ flat }(E,\mathbf{ k})$ into Dyson's equation
\begin{equation}
G_\mathrm{ terrace }(E,\mathbf{ k}) = ( E - H_\mathrm{ terrace}(\mathbf{ k}) - T^\dagger G_\mathrm{ flat }(E,\mathbf{ k}) T   )^{-1}
\end{equation}
which adds an incomplete atomic layer (described by $H_\mathrm{ terrace}$) on top of the sample ($T$ is the hopping matrix between the terrace and the last full layer), similarly to our previous work.\cite{Polley2018} The obvious drawback of this method is its computational complexity, as it requires calculating inverses of a very large, dense matrices, when an atomic terrace of the width of hundreds of lattice constants is considered. Furthermore, it allows only for calculation of periodic sequences of terraces -- such an approach requires defining as a periodic cell a row of half-infinite atomic columns terminating at the surface of the crystal. $G_\mathrm{ flat }$ of such a cell can be calculated quite efficiently, by utilizing the translation symmetry of the cell with a flat surface. Matrices $H_\mathrm{ terrace}$ and $T$ do not exhibit such symmetry .